# THE FRAMEWORK FOR IMPLEMENTING E-COMMERCE: THE ROLE OF BANK AND TELECOM IN BANGLADESH

Ijaj Md. Laisuzzaman, Nahid Imran, Abdullah Al Nahid, Md. Ziaul Amin, Md. Abdul Alim

**Abstract**— In this paper, we describe an effective framework for adapting electronic commerce or e-commerce services in developing countries like Bangladesh. The internet has opened up a new horizon for commerce, namely electronic commerce (e-commerce). It entails the use of the internet in the marketing, identification, payment and delivery of goods and services. At present internet facilities are available in Bangladesh. Slowly, but steadily these facilities are holding a strong position in every aspects of our life. E-commerce is one of those sectors which need more attention if we want to be a part of global business. Bangladesh is far-far away to adapt the main stream of e-commerce application. Though government is shouting to take the challenges of e-commerce, but they do not take the right step, that is why e-commerce dose not make any real contribution in our socio-economic life. Here we propose a model which may develop the e-commerce infrastructure of Bangladesh.

——————————  ◆  ——————————

## 1 INTRODUCTION

This is the era of information and communication technology. The leading concern of electronic revolution in this 21st century is to establish and ensure a better, easy and comfortable way of management, communication and development with the use of information technology. E-commerce has become a buzzword of present information technology. It is the process of conducting all forms of business through computer network and digital communication. Increasing domestic and global competition, economic downturn, rapidly changing market trends, and volatile financial markets have all added to the pressure on organizations to come up with effective responses to survive and succeed. Furthermore, easing of international trade barriers, economic liberalization, globalization, and deregulation have led to several challenges for organizations in developing and newly industrializing economies like Bangladesh.

The major principle of the paper is to provide an effective structure of implementing e-commerce in developing countries considering Bangladesh. In this paper we discussed about the role of bank and telecom sector for implementing e-commerce.

This paper is organized into eight sections. The basic concept on e-commerce is illustrated in section 2. In section 3, an overview of e-commerce infrastructure is presented. Section four is devoted to describe the emergence of the e-commerce in Bangladesh. The potential of implementing e-commerce is provided in section five. The growth of e-commerce in India and Sri Lanka are discussed in section six. In section seven, we proposed a framework for implementing e-commerce in Bangladesh based on current legislation of legal and regularities. This paper concludes with section eight.

## 2 E-COMMERCE

Commerce is an act of trade between two parties where the exchange is negotiated under a set of mutually acceptable conditions, so that both parties emerge satisfied with the result. The exchange may depend on whether the two parties are prepared to trust one another. E-commerce or electronic commerce is nothing but a business over the internet with the assistance of computers, which are linked to each other forming a network. To be specific, e-commerce would be buying and selling of goods and services and transfer of funds through digital communications. Most people think e-commerce means online shopping. But web shopping is only a small part of the picture. The term also refers to online stock, bond transactions, buying and downloading software without ever going to a store [1].

### 2.1 The Process of E-commerce

The process of e-commerce typically has the following phases [1]

1. Attract customers - This is usually done by providing pre-sales information on products and services. It may include on-line catalogues, price lists, product specifications etc.

2. Interact with customers - Interaction with customers is important. Customers are provided with all the information necessary before purchasing any goods. Negotiation with customers is also necessary. In this phase, a customer agrees with the terms of purchase.

3. Handle and manage orders - This phase fulfills the terms of the contract. Orders made by customers are captured. The exchange of payment and receipt and delivery logistics are done properly.

4. React to customers inquiries - It basically includes post-sales services comprising of new products information, product upgrade (e.g. software) etc. It can be used to maintain continuous contact with the customers.

## 2.2 E-commerce BusinesS Models

E-Commerce business models integrate the internet, digital communications and IT applications that enable the process of buying and selling. Models are [2]

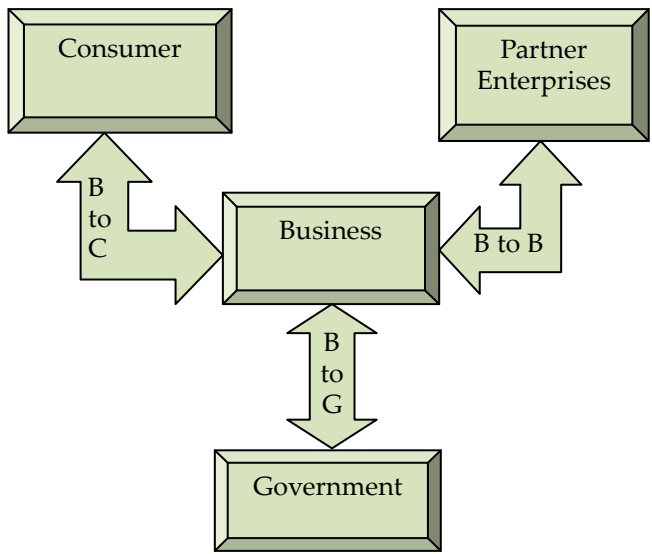

Fig. 1. Models of e-commerce

1. B2C (business to consumer)
2. B2B (business to business)
3. B2G (business to government)

Business to consumer: This is actually on-line shopping where the customers are provided with endless information via internet. In the websites, the customers search for desired goods by catalogs. The payment system is lot easier than conventional systems. On-line banking, on-line finance etc are important features of B2C. B2C e-commerce is unlikely to be of much use in the near future in Bangladesh because of low per capita income, a weak infrastructural and legal environment, lack of trust between business and consumers. In addition, non-availability of international credit cards, foreign currency remittance restrictions, delays and informal payments at customs clearance even for small value and quantity items will discourage B2C.

Business to business: It basically involves transaction between companies. B2B accounts for the majority of e-commerce. Here security plays a vital issue. The B2B application already exists in the export sector of Bangladesh, especially in the Ready Made Garments (RMG) industry. The RMG sector has begun to use the Internet, and its dependence on e-commerce is likely to grow in the coming years. The Internet would enable them to seek information about potential buyers as well as raw material suppliers. Similarly the practice of posting a website by individual producers has begun.

Business to government: It involves transaction between the government and any private company. B2G e-commerce is possible in Bangladesh, but on a limited scale at this stage. The government is a major buyer of goods and services from the private sector. Typically, the government procures goods and services by inviting tenders. The availability of the RFP (Request For Proposal) and other relevant documents on-line provides an alternate choice. Transactions involving information collection, obtaining various governmental forms, registering activities can also be conducted on-line.

## 2.3 Features of E-commerce

The features of e-commerce which make it considerably appreciable are as follows [3]:

1. Ubiquity: E-commerce is ubiquitous, meaning that it is available just about everywhere at all times. It liberates the market from being restricted to a physical space and makes it possible to shop from our desktop. From consumer point of view, ubiquity reduces transaction costs - the cost of participating in a market.

2. Global Reach: E-commerce technology permits commercial transactions to cross cultural and national boundaries far more conveniently and effectively as compared to traditional commerce. E-commerce can reach across geographic boundaries as well as demographic (e.g. age, income, gender, race, religion) boundaries.

3. Universal Standards: One prominent feature of e-commerce technologies is that the technical standards of the Internet and therefore the technical standards for conducting e-commerce are universal standards i.e. they are shared by all the nations around the world.

4. Interactivity: Unlike any of the commercial technologies of the twentieth century, with the possible exception of the telephone and mobile, e-commerce technologies are interactive. Consumers can interact with the content.

5. Information Density and Richness: The Internet vastly increases information density. It is the total amount and quality of information available to all market participants, consumers and merchants. E-commerce technologies reduce information collection, storage, communication and processing costs. Information richness refers to the complexity and content of a message.

6. Personalization: E-commerce technologies permit personalization. Merchants can target their marketing messages to specific individuals by adjusting the message to a person's name, interests and past purchases. The technology also permits customization.

## 2.4 E-commerce Architecture of an Organization

E-commerce is generally described as a method of buying and selling products and services electronically. The main vehicle of e-commerce remains the Internet and the World Wide Web, but uses of e-mail, fax and telephone orders are also prevalent. An organization's infrastructure for supporting an e-commerce service will typically have two main elements [1]:

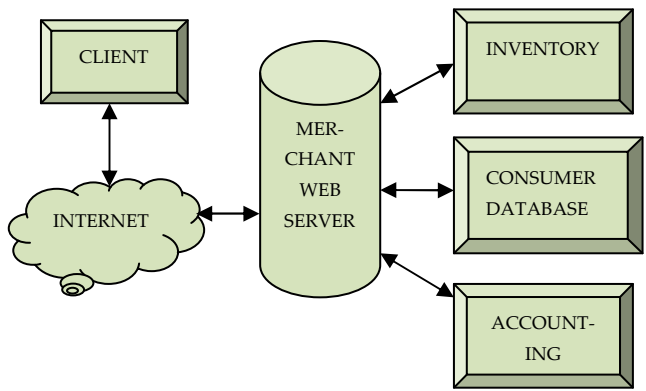

Fig. 2. E-commerce architecture

1. An e-commerce front-end (generally one or more web servers connected to the internet).
2. Back-end systems needed both to supply information to the front-end systems and to extract information from them.

## 3 INFRASTRUCTURE OF E-COMMERCE

Telecommunication network is major physical infrastructure for implementation of e-commerce. High speed, competitive international broadband access coupled with high density of local telecommunication facility is essential for growth of e-commerce. In developing country like Bangladesh telephone density, mobile phones, wireless communication, nationwide IP backbone etc. must be seriously considered for growth of this sector. Cyber cafes, kiosk, call center, etc. can be set up all over the country to provide some access to this technology at a reasonable cost while the infrastructure is being built.

It is important to create an environment on which consumers and sellers have confidence that is beneficial to e-commerce. For this, institutional background and policy changes may essential. The basic requirements of e-commerce are legal norms and standards covering contract enforcement, consumer protection, liability assignment, privacy protection, intellectual property rights. Process and technical standards regarding the way of payment are accepted in the Internet products are delivered to the consumers, security, authentication, encryption, digital signatures, connectivity protocols etc. The maximum business laws and regulations of Bangladesh are certain century old coming from subcontinent rules. Some laws are being updated. So the existing laws of commerce and evidence need to modify of the country.

Bangladesh government has made some efforts in establishing a regulatory framework for transactions over electronic media. A committee has been set up to draft cyber law about a year back but no appreciable progress has been made. Though there are no specific legal frameworks of international or national laws for e-commerce, additional laws have been enacted or are in the process in many countries. Moreover, UNCITRAL in 1996 has adopted a model law on e-commerce and some our neighboring countries like India has also enacted laws recognizing evidence and validity of electronic signature and contracts. This can be used as a basis for development of local laws and regulations [4].

## 4 APPEARANCE OF E-COMMERCE FOR BANGLADESH

Assume, a product is necessary for consumer which is not available in local area but available in other places. In that case information technology and e-commerce plays a vital role. Basically it happens in remote areas. It would be convenient for consumers if they could order online. It would save their time as well as money as consumers. For the country, it would enhance employment. Online transaction would increase the Gross Domestic Product (GDP) growth and thus help Bangladesh accomplish the Millennium Development Goals (MDGs). In the epoch of globalization, the Internet makes the world smaller and faster. E-commerce makes easy marketing and shopping from home. E-commerce makes possible business with customers over the internet. In e-commerce, customers can buy goods and services over the internet.

Bangladesh already has trained IT professionals. Its graduates are working with ISPs only Tk. 6000 (90 dollar!) per month. This shows that Bangladesh already has a plenty of trained IT professionals. Introduction of e-commerce would open new employment chances in the country [5].

The member of IT users in Bangladesh is increasing rapidly. According to International Telecommunication Union (ITU) report, Bangladesh had 450,000 internet users in 2007. In October 20, 2008, number of mobile phone users in Bangladesh stood at 45.09 million. More than 15,000 people are connected in "Facebook" through Bangladesh channels. All the districts headquarters have cyber cafes. Youth accounting for more than 35% of the total population gives Bangladesh an edge to choose e-commerce [5].

The government needs to make a realistic and time relevent business law. It must be worked in teamwork with the private sector to ensure a stable and reliable internet connection. E-commerce smooth the progress of payment of utility bills, fees etc.

## 5 IMPLEMENTATION OF E-COMMERCE IN BANGLADESH

In this era of globalization, e-commerce is a revolution in the field of trade and commerce. The rising of e-commerce has added a new dimension in the total scheme of business ensuring smarter and faster service in concerned sectors. The prime objective of e-commerce is to establish a user-friendly, trust-worthy, effective and efficient approach of providing internet centered business transactions. A lot of focus has been given on the technical factors for e-commerce like technical readiness assessment of Bangladesh for implementing e-commerce. Numbers of government and private organizations have accomplished several project based study for technical factor assessment. The achievement of e-commerce largely depends on the government along with "e". This overwhelming focus on "e" births the motivation to do the research on the non-technical environment of Bangladesh like political desire, vision and strategy, project management capacity etc. in respect to e-commerce implementation. There are some real lacks of evaluation of non-technical factors that contribute to adopt technology transfer. In [4], there is a case which is presented by Professor Richard Heeks is about the national databank project in Bangladesh. Planning commission of Bangladesh had taken an initiative to make official statistics available to government ministries, NGOs, and stakeholders for government and public use. Annual investment cost varies but only for network infrastructure it had spent US$440,000 during 1999/2000. However this project was total failure.

## 6 E-COMMERCE GROWTH OF OUR NEIGHBOUR COUNTRY INDIA AND SRI LANKA

The growth of electronic commerce has created the need for energetic and effective regulatory system, which would further reinforce the legal infrastructure that is vital to the success of electronic commerce. Most of India's banks and financial institutions have set up web sites. Online stock trading has also taken off in India with the Securities and Exchange Board of India (SEBI) making efforts to standardize message formats and address issues pertaining to technology, connectivity, security, surveillance and monitoring.

During the year 2000-2001, two major Industry Associations produced separate reports on e-commerce in India. Both the reports came out around the same time, namely June-July 2001. One was prepared by the National Committee on E-Commerce set up the Confederation of Indian Industry (CII), while the other was commissioned by the NASSCOM and prepared by the Boston Consulting Group. Both the reports are optimistic about the growth of e-commerce in India. The Confederation of Indian Industry (CII) report estimates the volume of e-commerce to grow to Rs 500 billion (US$ 10.6 billion) in the year 2003, out of which B2B will be Rs 420 billion (US$ 9 billion) and B2C will be Rs 80 billion (US$ 1.7 billion) (CII, 2001). The NASSCOM-BCG Report, on the other hand, estimates for the same year that the total volume of e-commerce will be Rs 1,950 billion (US$ 41.5 billion), out of which Rs 1,920 billion (US$ 41 billion) will be on account of B2B and Rs 3 billion (US$ 64 million) will be on account of B2C (NASSCON and BCG, 2001). E-commerce volume for the year is estimated to be Rs 150-200 billion (US$ 3.2-4.2 billion) (NASSCON and BCG, 2001) [7].

Late 2002 in response to the challenges of e-commerce Sri Lanka launched an initiative program named 'E-Sri Lanka Initiative: A Roadmap for the Strategic Vision for ICT development in Sri Lanka'. The E-Sri Lanka Initiative is a visionary program and an investment in the future of the Sri Lankan ICT sector. The government of Sri Lanka expects that the initiative will help address a problem faced by a number of developing countries in the South East Asia region: "We have the human resources, but it is converting that into real economic growth that is the challenge" (United World, 2002). In bellow figure 3 shows an outline of the strategic vision for E-Sri Lanka [8].

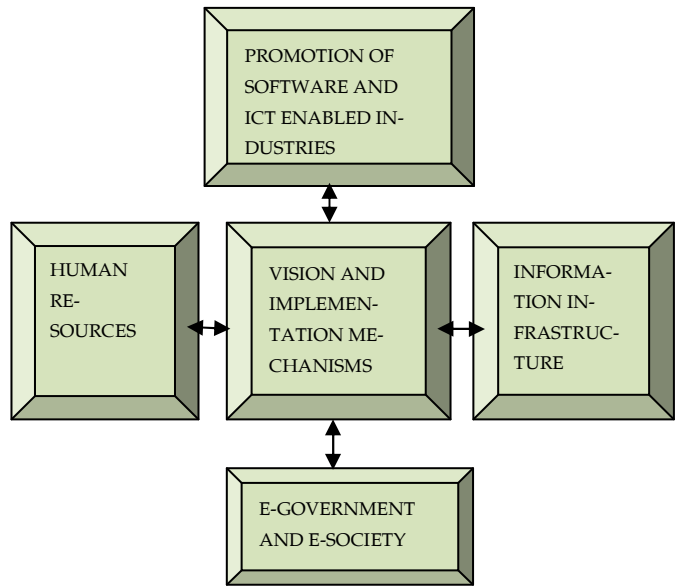

Fig. 3. Strategic ICT Roadmap: E-Sri Lanka Initiative

## 7 PROPOSED FRAMEWORK TO IMPLEMENT E-COMMERCE

In this section we proposed a framework to implement e-commerce in Bangladesh using banking and telecom support.

### 7.1 Banking Support for Implementing E-commerce

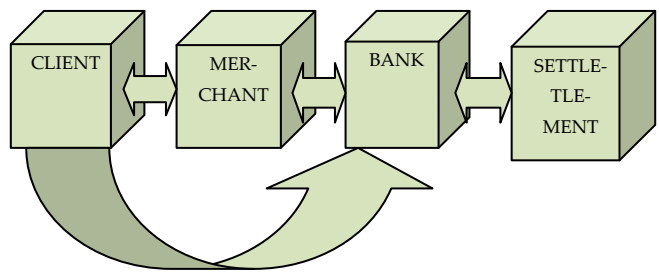

Fig. 4. Banking support for implementing electronic commerce

Finance and banking is an important sector for establishing e-commerce. Bank is the only authorized organi-zation which can store and transact money. Figure 4 shows the block diagram of the proposed framework of banking sector for implementing e-commerce. In this framework, client bank means the bank where client has an account and merchant bank means the bank where merchant has an account. For e-commerce, client/consumer and merchant need an account in any commercial bank which is capable of on-line transaction over internet. If client want to purchase any products from the merchant, he/she must need to brows the merchant website, choose appropriate product and fill up necessary requirements such as bank account number, credit card number, address etc. and also accept the rules and regulatory schemes. After this merchant send the information to his bank for processing the transaction. Then merchant send request to the client bank to transact money. But in the perspective of Bangladesh there are some regulatory problems. In Bangladesh inter bank on-line transaction is not allowed by central government bank "Bangladesh Bank". So, here we need a settlement bank which can process this transaction. The term "settlement bank" comes to overcome the regulatory problem. Any commercial bank can be a settlement bank. In this case every commercial bank of the country creates an account in the settlement bank as a user. Now, merchant bank send request to the settlement bank for processing the transaction. Then settlement bank draw money from the client bank account and deposit to the merchant bank account. Next, it send message to both banks. After this clearance client bank receive the money from client account and merchant bank deposit money to the merchant account.

There are some other roles of banking sector in e-commerce such as, online corporate banking, electronic fund transfer, automated teller machines (ATM), debit card, credit card etc.

There are some drawbacks of banking mechanism like distribution coverage, account opening, time consuming, and cost.

### 7.2 Arise of Telecom Sector to Overcome Drawbacks of Banking Mechanism

There are few drawbacks of banking sector in e-commerce such as, distribution coverage, banking account opening, time consuming, and cost. The distribution coverage of on-line connected commercial banks in Bangladesh is not very large. Opening an account is also a complicated process. It is very time consuming process to deposit money by maintaining serial. This also includes transport cost. So telecom plays a vital role in e-commerce to eliminate these drawbacks. In this case here telecom acts as a solution maker or payment gateway. In Bangladesh cellular phones are now available. Every cellular phone has a balance account and the phone number which are unique number. Distribution coverage of telecom operator in this country is very good. Any user can recharge his cell phone balance at any corner in the country. In this process a phone registration is necessary for security purpose. After registration operator give a unique security code to the user against his cell number. Here, cell number act as a bank account number and security code as a transaction ID instead of credit card number. So, in this method client is not necessary to open a bank account.

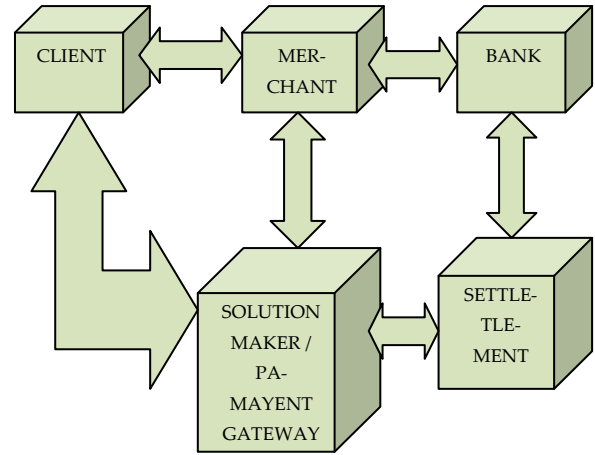

Fig. 5. The mechanism of telecom sector

Now when client purchase a product from the merchant, at first client fulfill the proper requirement. After this merchant send a request message to the solution maker or payment gateway to process this transaction. Solution maker also need an account in the settlement bank. Next, solution maker send request to the settlement bank for completing this transaction. Then settlement bank draw money from the solution maker account and deposit it to the merchant account. Next, it send message to the solution maker and merchant bank. After this clearance solution maker receive the money from clients cell phone balance account and merchant bank deposit money to the merchant account.

## 8 CONCLUSION

A developing country can become industrialized and modernized if it can extensively apply IT to enhance productivity and international competitiveness, develop e-commerce and e-governance applications. An information-based society or knowledge based society is composed of IT products, IT applications in society and economy as a whole. Many countries in Asia are taking advantage of e-commerce through opening of economies, which is essential for promoting competition and diffusion of Internet technologies. The Internet is boosting efficiency and enhancing market integration in developing countries. In this paper, we have presented an overview on e-commerce and its facilities. We also proposed a framework which is compatible current rules and regulations of Bangladesh. The implementation of e-commerce is the only line of attack to uplift the socio-economic infrastructure of the country into a glittering one.

## REFERENCES

[1] E-commerce Security, www.upu.int/security/en/e-commerce_security_en.pdf.

[2] Nazmul Hossain, E-commerce in Bangladesh: Status, Potential and Constraints, Draft Final Report, JOBS/IRIS Program of USAID, December 2000.
[3] Shailendra Sial, Unique Features of E-commerce, http://ezinearticles.com/?id=1887974
[4] Hasan Iqbal, "Economic Policy Paper on E-commerce a Business Link".
[5] AFM, Maniul Ahsan, MS Finance Student, Texas Tech University, http://www.thefinancialexpress-bd.com/2009/01/12/55733.html.
[6] Chowdhury Golam Hossan, Md. Wahidul Habib, I. Kushchu, Success and Failure Factors for e-Government projects implementation in developing countries: A study on the perception of government officials of Bangladesh.
[7] Mr. Rajiv Rastogi, India: Country Report on E-commerce Initiative, Country Presentation, Part Three.
[8] Michael S. Lane, Glen Van DerVyver, Sarath Delpachitra, Srecko Howard Lanem, vandervy, delpachi, howards@usq.edu.au, 'An Electronics Commerce Initiative in Regional Sri Lanka: The Vision for the Central Province Electronic Commerce Portal' Faculty of Business, University of Southern Queensland, Australia, EJISDC (2004) 16,1,1-18.
[9] Nitai Chandra Debnath, Abdullah Al Mahmud, The Environment of E-commerce in Bangladesh, Daffodil International University Journal of Business and Economics, Vol. 2, no. 2, July 2007.
[10] S. A. Ahsan Rajon, On the design and implementation of E-commerce in Bangladesh, Computer Science and Engineering Discipline, Khulna University, Khulna.
[11] Dr. Md. Abdul Mottalib, Sheikh Faridul Hasan, Syed Khairuzzaman Tanbeer, "BPLC: Proposed Approach to Implement E-governance in Remote Areas of Developing Countries", International Academy of CIO (IAC), Japan 2006.
[12] Journal of Engineering and Technology, Islamic University of Technology (IUT), The Organization of The Islamic Conference, Vol. 1, No. 1, January – June 2002.
[13] Michael Heng, Research Note: Implications of e-Commerce for Banking and Finance, No. 006/2006.
[14] Tanvir Ahmed, IT Revolution in Bangladesh: A Review
[15] Mohammad Mizanur Rahman, E-Banking in Bangladesh: Some Policy Implications.
[16] Shah Mohammad Sanaul Haque, Fatema-Tu-Zohra Binte Zaman, E-Government: Preparedness of Bangladesh Civil Service.

**Ijaj Md. Laisuzzaman**, Electronics and Communication Engineering Discipline, Khulna University, Khulna-9208, Bangladesh.

**Nahid Imran**, Electronics and Communication Engineering Discipline, Khulna University, Khulna-9208, Bangladesh.

**Abdullah Al Nahid,** Lecturer, Electronics and Communication Engineering Discipline, Khulna University, Khulna-9208, Bangladesh.

**Md. Ziaul Amin,** Lecturer, Electronics and Communication Engineering Discipline, Khulna University, Khulna-9208, Bangladesh.

**Md. Abdul Alim,** Assistant Professor, Electronics and Communication Engineering Discipline, Khulna University, Khulna-9208, Bangladesh.